\documentclass[sigconf, anonymous=false, balance=true, nonacm]{acmart}

\AtBeginDocument{%
  \providecommand\BibTeX{{%
    \normalfont B\kern-0.5em{\scshape i\kern-0.25em b}\kern-0.8em\TeX}}}

\usepackage{subcaption}

\usepackage{enumitem}
\setlist{nosep}
\usepackage[toc,acronym,nowarn]{glossaries}
\usepackage{todonotes}
\usepackage{listings}

\newacronym{OS}{OS}{Operating System}
\newacronym{SE}{SE}{Secure Element}
\newacronym{NFC}{NFC}{Near Field Communication}
\newacronym{BLE}{BLE}{Bluetooth Low Energy}
\newacronym{RIL}{RIL}{Radio Interface Layer}
\newacronym{SELinux}{SELinux}{Security-Enhanced Linux}
\newacronym{LSM}{LSM}{Linux Security Modules}
\newacronym{TCB}{TCB}{Trusted Computing Base}
\newacronym{MAC}{MAC}{Mandatory Access Control}
\newacronym{MMIO}{MMIO}{Memory-mapped I/O}
\newacronym{IPC}{IPC}{Inter-Process Communication}
\newacronym{ICC}{ICC}{Inter-Container Communication}
\newacronym{cgroups}{cgroups}{control groups}
\newacronym{CA}{CA}{Certificate Authority}
\newacronym{PKI}{PKI}{Public Key Infrastructure}
\newacronym{MITM}{MITM}{Man-In-The-Middle}
\newacronym{BYOD}{BYOD}{Bring-Your-Own-Device}
\newacronym{CM}{CM}{Container Management}
\newacronym{SM}{SM}{Security Management}
\newacronym{HAL}{HAL}{Hardware Abstraction Layer}
\newacronym{TLS}{TLS}{Transport Layer Security}
\newacronym{C2C}{C2C}{Container To Container}
\newacronym{Protobuf}{Protobuf}{Protocol Buffers}
\newacronym{SDO}{SDO}{Sensitive Data Object}
\newacronym{VMA}{VMA}{Virtual Memory Area}
\newacronym{PGD}{PGD}{Page Global Directory}
\newacronym{PUD}{PUD}{Page Upper Directory}
\newacronym{PMD}{PMD}{Page Middle Directory}
\newacronym[firstplural=Page Table Entries (PTEs)]{PTE}{PTE}{Page Table Entry}
\newacronym{COW}{COW}{Copy-On-Write}
\newacronym{IV}{IV}{Initialization Vector}
\newacronym{ESSIV}{ESSIV}{Encrypted Salt-Sector Initialization Vector}
\newacronym{KSM}{KSM}{Kernel Samepage Merging}
\newacronym{JIT}{JIT}{Just-In-Time}
\newacronym{DMA}{DMA}{Direct Memory Access}
\newacronym{FDE}{FDE}{Full Disk Encryption}
\newacronym{AS}{AS}{Address Space}
\newacronym{GCM}{GCM}{Google Cloud Messaging}
\newacronym{TPM}{TPM}{Trusted Platform Module}
\newacronym{JTAG}{JTAG}{Joint Test Action Group}
\newacronym{LUKS}{LUKS}{Linux Unified Key Setup}
\newacronym{VPN}{VPN}{Virtual Private Network}
\newacronym{PBKDF2}{PBKDF2}{Password-Based Key Derivation Function 2}
\newacronym{KVM}{KVM}{Kernel-based Virtual Machine}
\newacronym{VM}{VM}{Virtual Machine}
\newacronym{HV}{HV}{Hypervisor}
\newacronym{SEV}{SEV}{Secure Encrypted Virtualization}
\newacronym{SME}{SME}{Secure Memory Encryption}
\newacronym{TSME}{TSME}{Transparent \gls{SME}}
\newacronym{SP}{AMD-SP}{Secure Processor}
\newacronym[firstplural=Guest Virtual Adresses (GVAs)]{GVA}{GVA}{Guest Virtual Address}
\newacronym[firstplural=Guest Physical Addresses (GPAs)]{GPA}{GPA}{Guest Physical Address}
\newacronym[firstplural=Host Physical Addresses (HPAs)]{HPA}{HPA}{Host Physical Address}
\newacronym[firstplural=System Physical Addresses (SPAs)]{SPA}{SPA}{System Physical Address}
\newacronym{GPT}{GPT}{Guest Page Table}
\newacronym{HPT}{HPT}{Host Page Table}
\newacronym{TLB}{TLB}{Translation Lookaside Buffer}
\newacronym{PoC}{PoC}{Proof-of-Concept}
\newacronym{ORAM}{ORAM}{Oblivious RAM}
\newacronym{SEV-ES}{SEV-ES}{SEV Encrypted State}
\newacronym{SEV-SNP}{SEV-SNP}{SEV Secure Nested Paging}
\newacronym{RMP}{RMP}{Reverse Map Table}
\newacronym{VMCB}{VMCB}{Virtual Machine Control Block}
\newacronym{SLAT}{SLAT}{Second Level Address Translation}
\newacronym{SSH}{SSH}{Secure Shell}
\newacronym{RSA}{RSA}{Rivest–Shamir–Adleman}
\newacronym{ECDHE}{ECDHE}{Elliptic-Curve Diffie-Hellman Ephemeral}
\newacronym{AES}{AES}{Advanced Encryption Standard}
\newacronym{OOM}{OOM}{Out Of Memory}
\newacronym{MKTME}{MKTME}{Multi-Key Total Memory Encryption}
\newacronym{VMI}{VMI}{Virtual Machine Introspection}
\newacronym{MAD}{MAD}{Median Absolute Deviation}
\newacronym{HSM}{HSM}{Hardware Security Module}
\newacronym{AE}{AE}{Automatic Exit}
\newacronym{NAE}{NAE}{Non-Automatic Exit}
\newacronym{AES-NI}{AES-NI}{AES New Instructions}
\newacronym{NIC}{NIC}{Network Interface Card}
\newacronym{NMI}{NMI}{Non-Maskable Interrupt}
\newacronym{MTU}{MTU}{Maximum Transmission Unit}
\newacronym{VA}{VA}{Virtual Address}
\newacronym{GFN}{GFN}{Guest Frame Number}
\newacronym{SFN}{SFN}{System Frame Number}
\newacronym{IOMMU}{IOMMU}{I/O Memory Management Unit}
\newacronym{AISE}{AISE}{Address Independent Seed Encryption}
\newacronym{MT}{MT}{Merkle Tree}
\newacronym{BMT}{BMT}{Bonsai Merkle Tree}
\newacronym{LPID}{LPID}{Located Page IDentifier}
\newacronym{CB}{CB}{Counter Block}
\newacronym{PRD}{PRD}{Page Root Directory}
\newacronym{SWIOTLB}{SWIOTLB}{Software I/O Translation Buffer}
\newacronym{ASID}{ASID}{Address Space Identifier}
\newacronym{vCPU}{vCPU}{virtual CPU}
\newacronym{VC}{\texttt{\#VC}}{VMM Communication Exception}
\newacronym{GHCB}{GHCB}{Guest Hypervisor Communication Block}
\newacronym{IDT}{IDT}{Interrupt Descriptor Table}
\newacronym{KASLR}{KASLR}{Kernel Address Space Layout Randomization}
\newacronym{SLES}{SLES}{SUSE Linux Enterprise Server}
\newacronym{RHEL}{RHEL}{RedHat Enterprise Linux}
\newacronym{TSC}{TSC}{Time Stamp Counter}
\newacronym{IRET}{IRET}{Return from interrupt}
\newacronym{ROP}{ROP}{Return-oriented programming}
\newacronym{MSR}{MSR}{Model Specific Register}
\newacronym{RNG}{RNG}{Random Number Generator}
\newacronym{PRNG}{PRNG}{Pseudo Random Number Generator}
\newacronym{IBS}{IBS}{Instruction Based Sampling}
\newacronym{VMSA}{VMSA}{\gls{VM} Save Area}
\newacronym{OVMF}{OVMF}{Open Virtual Machine Firmware}
\newacronym{RW}{RW}{Read-Write}
\newacronym{RWX}{RWX}{Read-Write-Execute}

\makeglossaries
\glsdisablehyper

\definecolor{mGreen}{rgb}{1,0.6,0}
\definecolor{mGray}{rgb}{0.5,0.5,0.5}
\definecolor{mPurple}{rgb}{0.58,0,0.82}
\definecolor{backgroundColour}{rgb}{0.95,0.95,0.92}

\lstdefinestyle{CStyle}{
    backgroundcolor=\color{backgroundColour},
    commentstyle=\color{mGreen},
    keywordstyle=\color{magenta},
    numberstyle=\tiny\color{mGray},
    stringstyle=\color{mPurple},
    basicstyle=\footnotesize,
    breakatwhitespace=false,
    breaklines=true,
    captionpos=b,
    keepspaces=true,
    numbers=left,
    numbersep=5pt,
    showspaces=false,
    showstringspaces=false,
    showtabs=false,
    tabsize=2,
    language=C,
    xleftmargin=1em,
    framexleftmargin=1em
}

\renewcommand{\texttt}[1]{%
  \begingroup
  \ttfamily
  \begingroup\lccode`~=`/\lowercase{\endgroup\def~}{/\discretionary{}{}{}}%
  \begingroup\lccode`~=`-\lowercase{\endgroup\def~}{-\discretionary{}{}{}}%
  \begingroup\lccode`~=`\_\lowercase{\endgroup\def~}{\_\discretionary{}{}{}}%
  \catcode`/=\active\catcode`-=\active\catcode`\_=\active
  \scantokens{#1\noexpand}%
  \endgroup
}

\begin{document}

\title{Exploiting Interfaces of Secure Encrypted Virtual Machines}

\author{Martin Radev}
\email{martin.radev@aisec.fraunhofer.de}
\affiliation{%
  \institution{Fraunhofer AISEC}
  \city{Garching near Munich}
  \country{Germany}
}
\author{Mathias Morbitzer}
\email{mathias.morbitzer@aisec.fraunhofer.de}
\affiliation{%
  \institution{Fraunhofer AISEC}
  \city{Garching near Munich}
  \country{Germany}
}

\begin{abstract}
Cloud computing is a convenient model for processing data remotely.
However, users must trust their cloud provider with the confidentiality and integrity of the stored and processed data.
To increase the protection of virtual machines, AMD introduced SEV, a hardware feature which aims to protect code and data in a virtual machine.
This allows to store and process sensitive data in cloud environments without the need to trust the cloud provider or the underlying software.

However, the virtual machine still depends on the hypervisor for performing certain activities, such as the emulation of special CPU instructions, or the emulation of devices.
Yet, most code that runs in virtual machines was not written with an attacker model which considers the hypervisor as malicious.

In this work, we introduce a new class of attacks in which a malicious hypervisor manipulates external interfaces of an SEV or SEV-ES virtual machine to make it act against its own interests.
We start by showing how we can make use of virtual devices to extract encryption keys and secret data of a virtual machine.
We then show how we can reduce the entropy of probabilistic kernel defenses in the virtual machine by carefully manipulating the results of the \texttt{CPUID} and \texttt{RDTSC} instructions.
We continue by showing an approach for secret data exfiltration and code injection based on the forgery of MMIO regions over the VM's address space.
Finally, we show another attack which forces decryption of the VM's stack and uses Return Oriented Programming to execute arbitrary code inside the VM.

While our approach is also applicable to traditional virtualization environments, its severity significantly increases with the attacker model of SEV-ES, which aims to protect a virtual machine from a benign but vulnerable hypervisor.

\end{abstract}



\maketitle

\section{Introduction}

\glsresetall

Hardware virtualization is a common approach in cloud environments to securely and efficiently distribute hardware resources among multiple users.
Using virtualization, each user can create its own \gls{VM} to run any \gls{OS} and software.
Each \gls{VM} uses virtualized hardware, which is managed by the \gls{HV}.
The \gls{HV} has direct access to the \gls{VM}'s memory and architectural state.
This means that the \gls{HV} is always able to infer computations performed within the \gls{VM}.
Therefore, the user of the \gls{VM} has to fully trust the \gls{HV}, and hence the cloud provider.

In 2016, AMD released the \gls{SEV} feature to protect a \gls{VM} from a compromised \gls{HV}~\cite{kaplan2016amd}.
Since then, also the successors \gls{SEV-ES}~\cite{sev-es} and \gls{SEV-SNP}~\cite{AMD2020SNP} have been announced, which address additional attack vectors.
These improvements provide protection and confidentiality by encrypting the \gls{VM}'s memory and state, and limit the possibilities of targeted memory corruption by the \gls{HV}.
The \gls{SEV} features use a strong attacker model for which the confidentiality of the \gls{VM} must be guaranteed even if the attacker is in control of the \gls{HV}.

However, modern \glspl{OS}, which are also used inside the \glspl{VM}, have been developed over the years with the consideration that higher privileged layers are always trusted.
In comparison, \gls{SEV} introduces a new attacker model in which the \gls{HV} may be compromised and malicious~\cite{kaplan2016amd}.
This new attacker model requires to reevaluate the security constraints and to sanitize the no longer trusted interfaces between the \gls{VM} and \gls{HV}.
Such interfaces include, among others:
\begin{itemize}[nosep] 
\item the \texttt{virtio} interfaces for communicating data with external devices
\item instructions intercepted by the \gls{HV}
\item \glspl{MSR} for virtualization
\end{itemize}
These interfaces exist to provide correct and configurable virtualization, and to match a traditional attacker model in which the \gls{VM} is considered untrusted and the \gls{HV} needs to be protected.
However, with the attacker model of \gls{SEV}, these interfaces have to be considered untrusted and carefully validated from both sides.

As long as this sanitization of external interfaces is not performed, the \gls{VM} is potentially vulnerable to attacks from higher privilege layers.
We show that such attacks can be performed by a \gls{HV} on a \gls{VM} by adding cryptographic \texttt{virtio} devices to the \gls{VM}'s security components, or by providing semantically incorrect information such as CPU capabilities and time.
This allows a malicious \gls{HV} to force the \gls{VM} to act against its own interest, for example by 
1) exposing cryptographic keys and secret data to the \gls{HV}, 
2) reducing the entropy of its randomly generated values, 
3) leaking secrets or reading \gls{HV}-controlled data for most \texttt{MOV} memory accesses,  
4) decrypting its stack memory.
Even though such attacks can be applied on any virtualized environment, they are of much higher criticality on \gls{SEV}- and \gls{SEV-ES}-protected \glspl{VM} due to the different attacker model, which aims to protect the \gls{VM} from a vulnerable cloud provider~\cite{kaplan2016amd, singh2017x86}.

In summary, we make the following contributions: 

\begin{itemize}
    \item We introduce a technique which allows us to manipulate the \texttt{/dev/hwrng} randomness source in \glspl{VM} and to extract encryption keys from the Kernel Crypto API.
  \item We show how we are able to practically disable probabilistic kernel defenses in encrypted \glspl{VM} by intercepting \texttt{CPUID} and \texttt{RDTSC} instructions.
  \item We demonstrate how a malicious \gls{HV} can forge \gls{MMIO} regions to exfiltrate and inject data into encrypted \glspl{VM}.
  \item We present a fourth attack which allows us to execute code in encrypted \glspl{VM} by corrupting the \gls{VM}'s Guest Page Table.
\end{itemize}

\smallskip
Note that an early report, results and \glspl{PoC} were responsibly delivered to the AMD security team with which we discussed possible mitigations against the attacks.

\section{Background}

In this section, we give a quick introduction into hardware virtualization and AMD \gls{SEV} protection mechanisms.

\subsection{Hardware virtualization}\label{hw-virt}
Hardware virtualization allows to create virtual hardware configurations by using both software and hardware.
This technology enables the creation of multiple \glspl{VM} running on the same physical host.
Each \gls{VM} runs its own \gls{OS} and can have its own hardware configuration.
The \glspl{VM} are managed by the \gls{HV}, a combination of user-space and kernel-space code which performs the necessary operations to support the correct execution of the \gls{VM}.
Such operations include the allocation of physical memory for the \gls{VM}, providing CPU capability information or emulating special instructions.

In order to guarantee the security of the \gls{HV} and other \glspl{VM} on the same system, \glspl{VM} do not have complete access to the underlying hardware.
For security and correct emulation, the \gls{HV} emulates certain CPU instructions such as \texttt{IO instructions}, \texttt{MSR instructions}, \texttt{CPUID} and \texttt{RDTSC}.
The \texttt{CPUID} instruction returns information about CPU features.
If a \gls{VM} issues the \texttt{CPUID} instruction, the \gls{HV} is able to adapt the return values of the instruction.
This allows the \gls{HV} to disable certain CPU features for the \gls{VM} or to offer features which can be emulated.

The \texttt{RDTSC} instruction reads the CPU core's \gls{TSC}, a register which counts the number of cycles since the last reset~\cite{AMD2020}.
Each CPU core has its own \gls{TSC}, which is independent from the \glspl{TSC} on other cores.
To ensure that the different values of the \glspl{TSC} do not cause any problems when moving a \gls{VM} to a different core, the \gls{HV} is able to control the \gls{TSC} by means of interception or writing to relevant \glspl{MSR}.

\subsection{AMD SEV}
\label{sec:amd_sev}

In 2016, AMD introduced the \gls{SEV} feature~\cite{kaplan2016amd} to protect a \gls{VM} from a benign but vulnerable \gls{HV}.
\gls{SEV} achieves this goal by encrypting the \gls{VM}'s memory with an encryption key bound to the instance of the \gls{VM}.
This prevents the \gls{HV} from accessing the \gls{VM}'s memory in plaintext.

Not all of the \gls{VM}'s memory is encrypted because the \gls{VM} and the \gls{HV} need to exchange data such as network packets or disk drive blocks.
To facilitate this, the \gls{VM} can select which physical pages should be encrypted or unencrypted by setting or unsetting the \texttt{C-bit} in the \gls{GPT}~\cite{kaplan2016amd}.
During the boot stage of the \gls{VM}, the kernel marks almost all memory as encrypted and later marks memory regions used for data communication as unencrypted.

An important drawback of \gls{SEV} is that it does provide confidentiality to the \gls{VM}'s memory, but it does not guarantee its integrity or freshness.
Thus, \gls{SEV} is vulnerable to attacks which attempt to corrupt~\cite{du2017secure, li2019exploiting, wilke2020sevurity} or replay the \gls{VM}'s memory as well as memory remapping attacks~\cite{hetzelt2017security, morbitzer2018severed}.
Another problem with \gls{SEV} is that it allows the \gls{HV} to read and modify the \gls{VM}'s architectural registers during a \texttt{VMEXIT}.
These registers may contain secret information such as encryption keys and passwords~\cite{werner2019severest}, and should therefore not be exposed to the untrusted \gls{HV}.

\begin{figure}[htbp]
    \begin{center}
    \includegraphics[width=1.0\columnwidth]{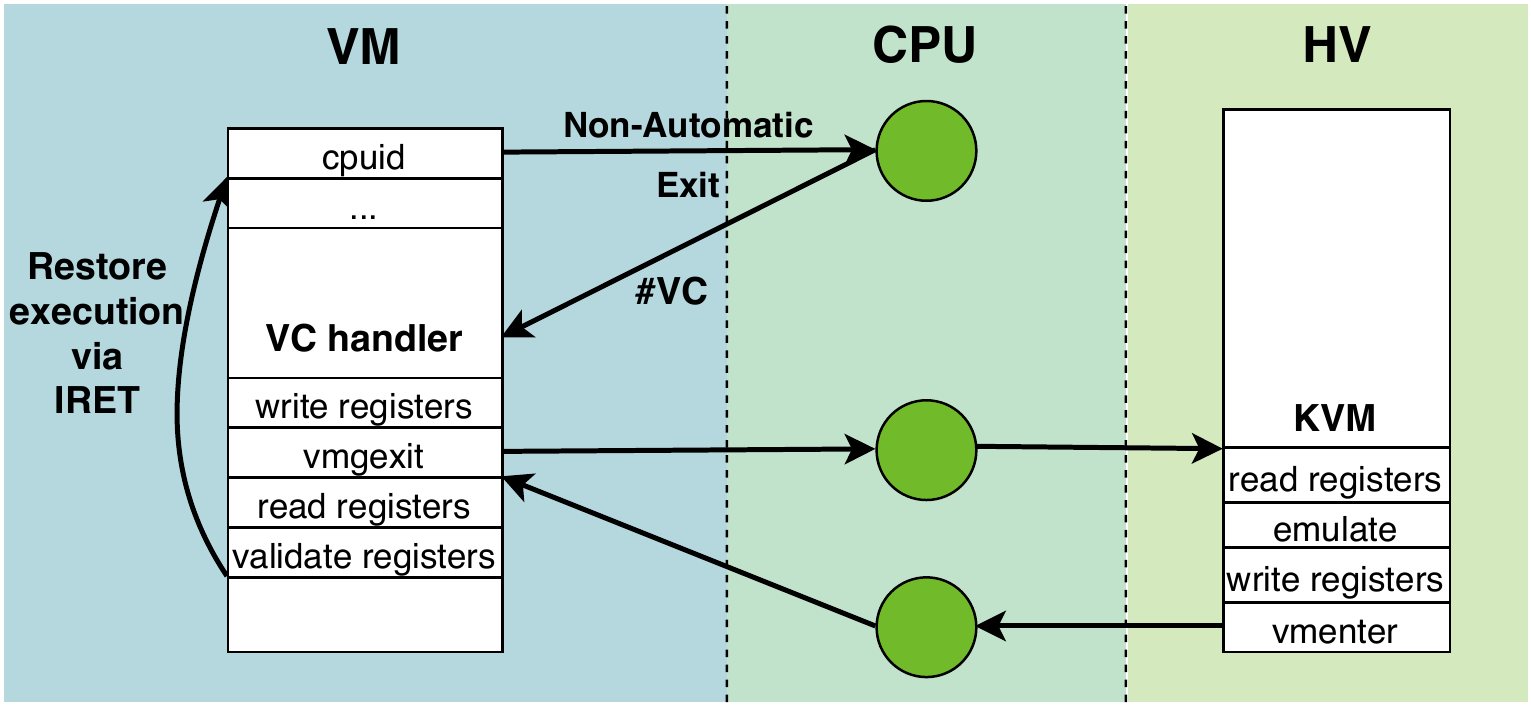}
    \end{center}
    \caption{On a Non-Automatic exit, the \gls{VM} first passes control to its VC handler.
                  Only afterwards, the VC handler gives the control to the \gls{HV}.
                  After resuming the \gls{VM}, the VC handler validates the information from the \gls{HV} before returning to the original instruction.}
    \label{fig:vc}
\end{figure}

To reduce the risk of exposing architectural registers to the \gls{HV}, AMD announced \gls{SEV-ES} in 2017~\cite{sev-es}.
\gls{SEV-ES} encrypts and integrity protects the architectural state of the \gls{VM} on a \texttt{VMEXIT}.
However, the \gls{HV} still needs to be able to read and modify some of the \gls{VM}'s registers in order to emulate intercepted instructions such as \texttt{CPUID} or \texttt{RDTSC}.
\gls{SEV-ES} addresses this issue by introducing a new unencrypted region, called the \gls{GHCB}.
The \gls{GHCB} contains only register values selected by the \gls{VM}, and is accessible by both the \gls{HV} and the \gls{VM}.
This method allows for communication of register values.

Figure~\ref{fig:vc} shows a sequence diagram for executing an intercepted instruction.
When the \gls{HV} intercepts an instruction, the CPU performs a \gls{NAE}.
This causes a \gls{VC}, which gets handled by the \gls{VM}'s VC handler.
The VC handler determines the cause of the exception and decides whether to invoke the \gls{HV} to emulate the instruction.
If the \gls{HV} needs to be invoked, the \gls{VM} copies the necessary registers into the \gls{GHCB} and executes the \texttt{VMGEXIT} instruction to invoke the \gls{HV}.

In case of intercepting for example the \texttt{CPUID} instruction, the \gls{VM} stores the values of the \texttt{EAX} and \texttt{ECX} registers into the \gls{GHCB} and executes a \texttt{VMGEXIT}.
When the \gls{HV} receives the \texttt{VMGEXIT}, it first reads the \texttt{EAX} and \texttt{ECX} registers.
Afterwards, the \gls{HV} emulates the \texttt{CPUID} instruction and writes the new values of the \texttt{EAX}, \texttt{ECX}, \texttt{EBX}, \texttt{EDX} registers into the \gls{GHCB}.
To finalize, the \gls{HV} hands control back to the \gls{VM} by issuing a \texttt{VMENTER}.
Once the \gls{VM} continues execution, it can validate the registers passed by the \gls{HV} and decide whether to update the architectural state.
Then, the \gls{VM} issues a \gls{IRET} instruction to continue execution immediately after the instruction which was originally intercepted.

In 2020, AMD announced an update to \gls{SEV-ES} named \gls{SEV-SNP}~\cite{AMD2020SNP}.
\gls{SEV-SNP} further enhances protection by addressing attacks based on memory corruption and page remapping.
\gls{SEV-SNP} also introduces a \texttt{CPUID} reporting feature which allows the \gls{VM} to verify the available CPU capability features with the firmware~\cite{amd2020snpabi}.
However, this feature does not prevent the \gls{HV} from disabling certain CPU capabilities for the \gls{VM}.
As \gls{SEV-SNP} was only recently standardized, we are not aware of any AMD CPU that supports \gls{SEV-SNP} at the time of writing.

\section{Unsafe virtio devices}\label{sec:steal}

In most virtualized environments, the Linux kernel makes use of the \texttt{virtio} interface~\cite{oasis2019virtio} to efficiently communicate data to external devices such as network cards and disk drives.
To establish communication via \texttt{virtio}, the \gls{VM} queries the identifiers and configurations of the available virtual devices from the \gls{HV}.
Afterwards, the \gls{VM}'s kernel automatically loads the corresponding device drivers.
For example, when the \gls{VM} attempts to write to disk or to send a network packet, the corresponding \texttt{virtio} driver in the \gls{VM} would consume the request.
Afterwards, it writes the data to a suitable guest physical address and signals the \gls{HV} to propagate the request to the physical hardware device.
The added layers of abstraction benefits performance because the \texttt{virtio} communication layers are aware of the virtualized environment and can transfer data more efficiently by reducing the number of \gls{VM} exits.

The improved performance has led to additional \texttt{IO} devices making use of \texttt{virtio}, including devices essential for the \gls{VM}'s security.
In this section, we show how two of those devices --- \texttt{virtio-rng}~\cite{qemu-rng} and \texttt{virtio-crypto}~\cite{qemu-crypto} --- pose a security risk when considering \gls{SEV}'s attacker model.

The \texttt{virtio-rng} device is a \gls{RNG} device, which allows to add entropy to the kernel's entropy pool~\cite{qemu-rng}.
After initialization, the device is accessible within the \gls{VM} via the \texttt{/dev/hwrng} interface.
By default, \gls{SEV}-protected \glspl{VM} do initialize the \texttt{virtio-rng} device~\cite{amdsevgit}.
However, considering \gls{SEV}'s attacker model, a device controlled by an untrusted \gls{HV} providing entropy to the \gls{VM} represents a critical attack vector.
For example, entropy-reliant software in the \gls{VM} might utilize the \texttt{/dev/hwrng} interface when seeding an \gls{RNG} for cryptographic operations.
This would allow the \gls{HV} to fully control the entropy provided to the software in the \gls{VM}.

Another device which represents a critical attack vector when considering \gls{SEV}'s attacker model is \texttt{virtio-crypto}~\cite{qemu-crypto}.
This device allows a \gls{VM} to make use of hardware features to accelerate cryptographic transformations.
Using the device, the \gls{VM} utilizes the \texttt{virtio} interface to deliver the cryptographic keys and data to the \gls{HV}.
After processing the information in hardware or software, the \gls{HV} returns the information to the \texttt{virtio-crypto} driver in the \gls{VM}.
This process allows the \gls{HV} to extract cryptographic information processed by the \gls{VM}.

In order for the \gls{VM} to use the \texttt{virtio-crypto} engine, a malicious \gls{HV} needs to
$(i)$ avoid that the \gls{VM} uses the default \texttt{aesni\_intel} engine and $(ii)$ ensure that the \texttt{virtio-crypto} device is registered for the \gls{VM}. 
By announcing the \texttt{AES-NI} CPU extension as unavailable when the \gls{VM} issues a \texttt{CPUID} instruction, we are able to achieve the former.
As the \gls{HV} is also in charge of launching the \gls{VM}, we are able to achieve the latter by registering the \texttt{virtio-crypto} device when launching the \gls{VM}.
This approach allows us to trick the \gls{VM} into using the \texttt{virtio-crypto} engine, allowing us to extract secret cryptographic information passed to the \gls{VM}'s Kernel Crypto API.

\section{Controlling the entropy sources}\label{sec:control}

In this section, we show how the Linux kernel uses different sources of entropy for its probabilistic defenses.
Afterwards, we demonstrate how a malicious \gls{HV} is able to reduce these sources of entropy in order to have the \gls{VM} use deterministic values for its presumably probabilistic defenses.

\subsection{Probabilistic kernel defenses}
\label{subsec:kdef}

To protect itself against attacks, the Linux kernel makes use of various mechanisms such as \gls{KASLR} and stack canaries~\cite{kernel2020self}.

\gls{KASLR} complicates exploitation of vulnerabilities such as buffer overflows in the kernel.
It achieves this by randomizing the virtual and physical offsets of the kernel image at boot stage.
Due to the randomized offsets, it is difficult to utilize techniques such as \gls{ROP}~\cite{roemer2012return}.
The randomization offered by \gls{KASLR} has at most nine bits of entropy for the virtual offset and the limit for the physical offset is determined by the size of available physical memory.

\gls{KASLR} also makes it difficult to directly read or write any physical memory address and to exploit the kernel's heap by also randomizing other memory regions such as the direct-physical map (\texttt{page\_offset\_base}).
The \emph{direct-physical map} determines the virtual address of the whole physical memory mapped into the kernel's address space.
Additional randomizations include the \emph{vmalloc} area (\texttt{vmalloc\_base}), which determines the address of the kernel's \emph{vmalloc heap}, and
the \emph{virtual-memory map} (\texttt{vmemmap\_base}), which determines the base address of the data structure holding the meta information for all physical pages.

Another probabilistic defense are \emph{stack canaries}, which help to detect stack buffer overflows.
A stack canary is a random token that is placed at the top of the stack frame between variables and saved special-purpose registers such as the \texttt{RIP} and \texttt{RBP}.
A modified stack canary indicates that special purpose registers could have been overwritten.
To verify the stack canary, the compiler adds code to the function's epilogue to check if the stack canary is corrupted.
Overwriting the stack canary with the same value is difficult for an attacker, as the canary offers up to 56 bits of entropy.
Although guessing the randomized values for KASLR and stack canaries is possible, a single wrong guess would be detected and require a system reboot.

\subsection{Early kernel entropy generation}\label{subsec:kent}

Common probabilistic kernel defenses (Section~\ref{subsec:kdef}) rely on randomizing different parameters to make them unpredictable for an attacker.
To ensure that the values are unpredictable, the kernel requires a reliable source of entropy.
The two most common sources of entropy during the early boot stage under the x86 architecture are \texttt{RDRAND} and \texttt{RDTSC}, which are also used by the Linux kernel.

\begin{figure}[htbp]
    \begin{lstlisting}[style=CStyle]
unsigned long kaslr_get_random_long(...) {
    unsigned long raw, random = get_boot_seed();
    ...
    if (has_cpuflag(X86_FEATURE_RDRAND))
        if (rdrand_long(&raw))
            random ^= raw;
    if (has_cpuflag(X86_FEATURE_TSC))
        random ^= rdtsc();
    ...
    return random;
}\end{lstlisting}
    \caption{The \texttt{kaslr\_get\_random\_long} function used to calculate \gls{KASLR} offsets.
             While Line $2$ creates an initial random state, Lines $4$ and $7$ make use of \texttt{RDRAND} and \texttt{RDTSC} to improve random number generation.}
    \label{code:get_long}
\end{figure}

During the boot process, the early Linux boot code has to compute the \gls{KASLR} physical and virtual offsets before decompressing the kernel image.
Both offsets are computed by the function \texttt{kaslr\_get\_random\_long} shown in Figure~\ref{code:get_long}.
In Line 2, the initial random value is determined by the \texttt{get\_boot\_seed} function which returns a hash of the kernel's build string and the boot parameters structure.
The boot parameters structure is allocated, zeroed-out and populated by the OS loader earlier in the boot stage as dictated by the Linux boot protocol~\cite{linux-boot-protocol}.
Starting with Line $4$, if supported by the CPU, the function makes use of \texttt{RDRAND} to add entropy to the random value.
Additionally, starting with Line $7$, if supported by the CPU, \texttt{RDTSC} is also used to add entropy to the random value.

\begin{figure}[htbp]
    \begin{lstlisting}[style=CStyle]
void __init kernel_randomize_memory(void) {
  ...
  prandom_seed_state(&rand_state,
    kaslr_get_random_long());
  ...
  for (i=0; i < ARRAY_SIZE(kaslr_regions);i++) {
    prandom_bytes_state(&rand_state, &rand, 8U);
    entropy = (rand % (entropy + 1)) & P4D_MASK;
    vaddr += entropy;
    *kaslr_regions[i].base = vaddr;
  }
}\end{lstlisting}
    \caption{The \texttt{kernel\_randomize\_memory} function used for randomizing the memory regions.
             To initialize the Pseudo Random Number Generator in Line $3$, the function makes use of the \texttt{kaslr\_get\_random\_long} function.}
    \label{code:mem_regions}
\end{figure}

Afterwards in the initialization phase, the kernel computes random offsets for the memory regions \texttt{vmalloc\_base},  \texttt{vmemmap\_base} and \texttt{page\_offset\_base}.
Figure~\ref{code:mem_regions} shows the respective function \texttt{kernel\_randomize\_memory}.
In Line $3$, the function initializes the \gls{PRNG} state with the value returned from the function \texttt{kaslr\_get\_random\_long} we just saw.

\begin{figure}[htbp]
    \begin{lstlisting}[style=CStyle]
void add_device_randomness(const void *buf,
        unsigned int size) {
    u64 time = random_get_entropy() ^ jiffies;
    ...
    _mix_pool_bytes(&input_pool, &time, 8U);
    ...
}\end{lstlisting}
    \caption{The \texttt{add\_device\_randomness} function used for adding entropy to the Linux entropy pool.
             The value of \texttt{random\_get\_entropy()} in Line $3$ corresponds to the value of \texttt{RDTSC}, and the value of the \texttt{jiffies} variable remains constant in the early kernel initialization phase.}
    \label{code:add_rand}
\end{figure}

Later in the kernel initialization stage, instantiated device drivers can add entropy to the Linux entropy pool by calling the function \texttt{add\_device\_randomness}.
Figure~\ref{code:add_rand} shows the relevant code of this function.
In Line $3$, the return value of \texttt{random\_get\_entropy} equals the return value of \texttt{RDTSC}, and the number of jiffies is constant at this stage since no timer interrupts have been yet received.
Afterwards in Line $5$, the function mixes the entropy into the entropy pool.
During the Kernel's lifespan, entropy is also added to the entropy pool by other events such as interrupts, user input events, and disk usage~\cite{muller2019documentation}.
The corresponding functions also use \texttt{RDTSC} as part of the random data mixed into the entropy pool.

Similarly to \gls{KASLR}, stack canaries of kernel threads and userspace processes receive random bytes from the Linux \gls{RNG}, which in turn receives its entropy from the entropy pool.
Therefore, as long as the entropy injected into the pool is known to the \gls{HV}, the entropy of any stack canary is significantly reduced.

\subsection{Entropy manipulation} \label{section:entropy_man}

We have now seen that many of the probabilistic kernel defenses rely on \texttt{RDRAND} and \texttt{RDTSC} as sources of entropy.
Thus, these defenses are only as good as the results returned from these instructions.
Next, we show how a malicious \gls{HV} is able to manipulate the return values of these instructions to eliminate the dependent probabilistic kernel defenses.

The first behavior we are using to our advantage is the fact that the \gls{HV} controls the CPU capabilities offered to the \gls{VM}.
Having control over the \gls{HV}, we can inform the \gls{VM} that \texttt{RDRAND} is unsupported by the CPU.
This prevents the \gls{VM} from using its main source of entropy during the early boot and kernel initialization stage.

Instead, the \gls{VM} and therefore all its probabilistic kernel defenses now only rely on \texttt{RDTSC} as a single source of entropy during the boot and kernel initialization stages.
Additionally, we are able to intercept the  \texttt{RDTSC} instruction, which allows us to arbitrarily adjust the return value of \texttt{RDTSC}.
This results in full control over the \gls{VM}'s only source of entropy during the early stage of the kernel.
Therefore, we are able to manipulate the probabilistic kernel defenses of the \gls{VM} and to make them deterministic.

The only aspect we have to consider when intercepting and emulating \texttt{RDTSC} is that the \gls{VM} also uses \texttt{RDTSC} for adding cycle delays and calibrating timer interrupts.
Therefore, improper emulation would resolve in hanging kernel threads. 
Thus, we have to differ between \texttt{RDTSC} being executed for entropy and for a different usage.

With the first version of \gls{SEV}, it is trivial to identify why the \gls{VM} is calling \texttt{RDTSC} since we are able to read the \texttt{RIP} register of each of the \gls{VM}'s \texttt{vCPUs}.
By statically inspecting the \gls{VM}'s kernel image, we are able to infer the location of the kernel and to select which \texttt{RDTSC} executions to manipulate.

Using \gls{SEV-ES} or \gls{SEV-SNP}, the \gls{VM}'s registers are encrypted, which prevents us from determining the \gls{VM}'s \texttt{RIP}.
However, we made the observation that different \texttt{RDTSC} instructions are typically used in functions which are located on separate pages in the \gls{VM}'s physical memory.
Furthermore, the function calls which precede the one using \texttt{RDTSC} are also typically located on different pages.
This allows us to identify the \texttt{RDTSC} usage by using the sequence of recent page accesses as context for the prediction.

We determine the most recent page accesses by manipulating the \gls{SLAT} table of the \gls{VM}.
Such manipulation cannot be performed solely with the Linux virtualization solution \texttt{KVM}, as it does not provide sufficient control over the \gls{SLAT} table.
Instead, we make use of the \textit{SEVered framework}~\cite{severedframework} to monitor page accesses.
The SEVered framework extends \texttt{KVM} and allows to remove the \texttt{present} flag from the \gls{VM}'s physical pages.
As soon as the \gls{VM} attempts to access any data on such a page, the resulting page fault causes the \gls{VM} to trap into the \gls{HV}.
This allows the \gls{HV} to track the different memory accesses of an \gls{SEV}(-ES)-protected \gls{VM} at page granularity.

We make use of this possibility to manipulate the entropy of a \gls{VM} at boot.
For this, we utilize the fact that the sequence of page faults before decompressing the kernel is deterministic, as already briefly mentioned by Wilke et al.~\cite{wilke2020sevurity}.
This behavior allows us to easily identify the page access after which the \gls{VM} twice executes \texttt{RDTSC}.
While the \gls{VM} uses the first execution to determine the physical \gls{KASLR} offset, the second execution determines the virtual offset.
By intercepting both executions and specifying a fixed return value, we are able to pin the physical as well as the virtual \gls{KASLR} offset.

The next page fault indicates the location in the \gls{VM}'s physical memory to which the \gls{VM} decompresses the kernel image.
This allows us to calculate the locations of all \gls{VM} kernel functions in physical memory by statically inspecting the \gls{VM}'s kernel image.

By analyzing the \gls{VM}'s page accesses at boot time, we found suitable page fault sequences which uniquely identify the usages of \texttt{RDTSC} as a source of entropy.
We modified \texttt{KVM} to include a state machine to easily allow us to specify what actions to perform when a page fault occurs or an \texttt{RDTSC} is intercepted.
This allowed us to always detect when the \gls{VM} uses \texttt{RDTSC} as a source of entropy in order to pass a known fixed value to the \gls{VM}.

\begin{figure}[t]
    \begin{center}
        \includegraphics[width=1.\columnwidth]{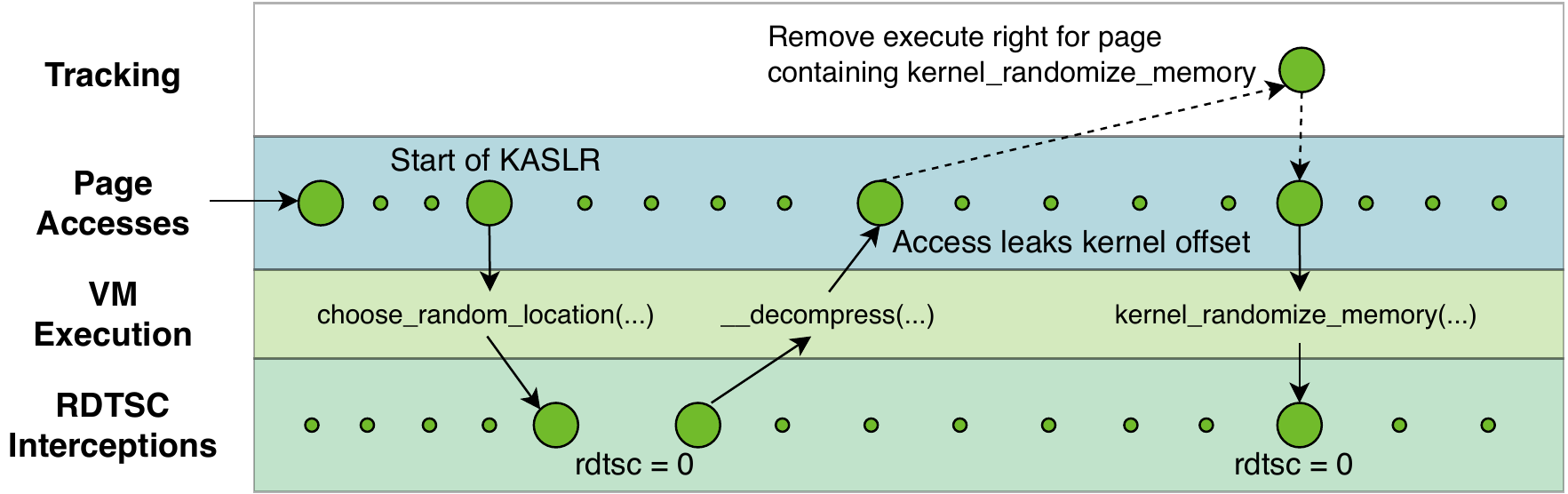}
    \end{center}
    \caption{Monitoring the \gls{VM} page access not only allows us to set the return value of the \texttt{RDTSC} instruction to zero at the required point in time, but also allows us to determine the \gls{KASLR} physical offset.
            }
    \label{fig:time_seq}
\end{figure}

Figure~\ref{fig:time_seq} shows the sequence of the steps we perform to manipulate the virtual and physical \gls{KASLR} offset.
The tracking lane shows which pages we track, and the page accesses lane shows the physical pages accessed by the \gls{VM}.
The \gls{VM} execution lane shows the execution flow in the \gls{VM}, and the \texttt{RDTSC} interceptions lane shows which return values we set to zero.
The \gls{HV} tracks the naturally occurring page faults until the \gls{VM} accesses the page containing the function responsible for selecting random locations for the kernel.
We then set the return values of the following two \texttt{RDTSC} executions to zero to pin the \gls{KASLR} physical and virtual image offsets.
The following page access is caused by the early Linux boot code decompressing the kernel image into memory, and discloses the \gls{KASLR} physical offset.
This allows us to determine the location of the \texttt{kernel\_randomize\_memory} function within the \gls{VM}'s memory.
We use this knowledge to track the respective page by removing its \texttt{execute} permission.
Once the \gls{VM} accesses the page, we set the return value of the subsequent \texttt{RDTSC} execution to zero.
While only requiring a single interference with the \gls{SLAT} table, these steps allow us to pin the \gls{KASLR} code and memory regions offsets between reboots of the \gls{VM}.

\subsection{Stealthy manipulation of the TSC} \label{section:tscmsr}

Additional difficulties for an attacker arise under \gls{SEV-ES} and \gls{SEV-SNP} when attempting to intercept the \texttt{RDTSC} instruction.
Intercepting the instruction is required in order to manipulate its return value (Section~\ref{section:entropy_man}).
However, when using \gls{SEV-ES} or \gls{SEV-SNP}, such an interception can be detected by the \gls{VM} as it causes a \texttt{VC} exception.
The \gls{VM} captures the exception and then performs a \texttt{VMGEXIT} to request the \gls{HV} to emulate the instruction.
After the \gls{HV} finished the emulation, the \gls{VM} can sanitize the returned values.

This approach allows the \gls{VM} to detect an unexpected \texttt{RDTSC} interception and manipulation of the return value by the \gls{HV}.
A \gls{VM} aware of this attack vector might forbid interception of \texttt{RDTSC} and simply halt execution.

However, the AMD Open-Source Register Reference~\cite{amd2018regman} also discusses other possibilities for the \gls{HV} to manipulate the \gls{VM}'s \texttt{TSC}.
These exist to ensure correct execution of the \gls{VM} and smoothless relocation to a different core or to other CPUs where the \texttt{TSC} might have a different value.
The additional possibilities include the \texttt{TSC Ratio MSR} and the \texttt{TSC Offset} field in the Virtual Machine Control Block structure.
The \texttt{TSC Ratio MSR} allows the \gls{HV} to provide a fixed-point multiplier for the elapsed time. 
Additionally, the \texttt{TSC Offset} field is an offset added to the value of the \texttt{TSC} when the \gls{VM} reads it.
Setting the value of the \texttt{TSC Ratio MSR} and the value of the \texttt{TSC Offset} to zero will effectively zero-out the \texttt{TSC} when read by the \gls{VM}.
Therefore, using the \texttt{TSC Ratio MSR} and the \texttt{TSC Offset} allows us to specify any value for the \texttt{TSC} without intercepting the \texttt{RDTSC} instruction.
For the rest of this paper, we assume the \texttt{TSC Offset} to be set to zero.

This approach requires careful calibration of the attack.
Unlike \texttt{RDTSC} interception, manipulation of the \texttt{TSC Ratio MSR} or \texttt{TSC Offset} does not allow us to determine how many times the \gls{VM} calls \texttt{RDTSC}.
Instead, we need to select a period of time during which the TSC will be zero.
This period must not include calls of the \gls{VM} to \texttt{RDTSC} for which the return value zero could cause problems for the \gls{VM}.
Examples for such calls are when \texttt{RDTSC} is used for adding delay or calibrating timer interrupts.
Also, we need to consider that the \texttt{TSC Ratio MSR} is a register available on each core.
If the \gls{VM} is rescheduled to execute on another core, we must ensure that the \texttt{TSC Ratio MSR} value on the respective cores are updated.

\begin{figure}[t]
    \begin{center}
        \includegraphics[width=\columnwidth]{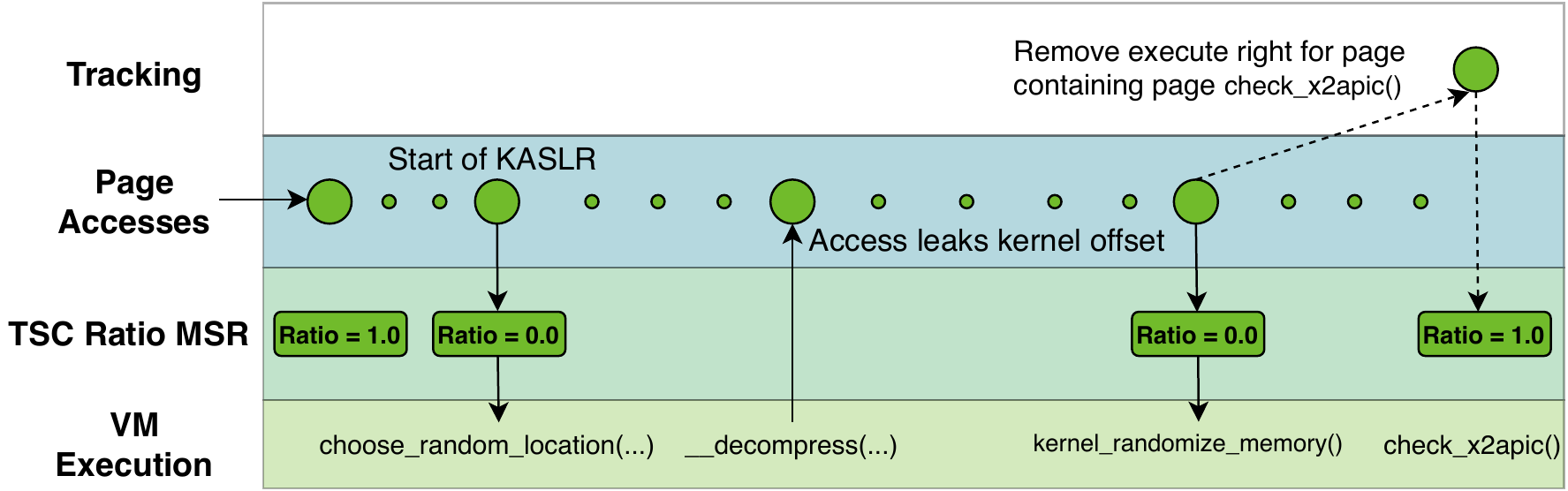}
    \end{center}
    \caption{Diagram for TSC Ratio MSR manipulation.
              The \texttt{TSC Ratio MSR} is kept at $0.0$ starting at the call to \texttt{choose\_random\_location} until after the call to \texttt{kernel\_randomize\_memory}.}
    \label{fig:time_seq_ratio}
\end{figure}

Figure~\ref{fig:time_seq_ratio} shows how we use the \texttt{TSC Ratio MSR} to manipulate the \gls{VM}'s early entropy sources.
The tracking, page accesses and VM execution lanes serve the same function as before (Section~\ref{section:entropy_man}),
and the \texttt{TSC Ratio MSR} lane shows the current value of the register.
Once the \gls{VM} starts, the \texttt{TSC Ratio MSR} has the default value of \texttt{1.0}.
When the \gls{VM} first accesses the page containing \texttt{choose\_random\_location}, we set the value of the \texttt{TSC Ratio MSR} to \texttt{0.0}.
This returns the value \texttt{0*<TSC-value>} to the \gls{VM}'s \texttt{RDTSC} instruction.
Subsequently, the \gls{VM} will start to decompress its kernel image.
Therefore, the following page access will leak the \gls{VM}'s KASLR physical offset which discloses the physical locations of all kernel functions of the \gls{VM}.
At this point, we further keep the \texttt{TSC Ratio MSR} value at \texttt{0.0} until \texttt{kernel\_randomize\_memory} returns.
By statically examining the Linux image, we determined that the kernel afterwards calls the function \texttt{check\_x2apic}, which is located on a different page.
This allows us to use the respective page as a trigger.
By removing the execute permission of the page, we are able to determine when the kernel calls the \texttt{check\_x2apic} function.
Once the \gls{VM} attempts to fetch an instruction from this page, we set the \texttt{TSC Ratio MSR} back to its default value and resume the \gls{VM}.

Using the provided steps, we are able to pin the KASLR code and memory regions offsets between reboots of the \gls{VM}.
This approach additionally complicates detection from within the \gls{VM}, as this attack does not require interception of the \texttt{RDTSC} instruction.

\section{Data Exfiltration and Injection via MMIO region forgery}
\label{sec:mmio_npf}

In the previous section, we showed how we are able to simplify locating secrets within an encrypted \gls{VM} by manipulating its sources of entropy.
In this section, we show how located secrets can afterwards be exfiltrated, or malicious data be injected, by forging \gls{MMIO} regions in the \gls{VM}'s address space under \gls{SEV-ES}.

A \gls{VM}'s software can program devices either by using I/O instructions, or by performing regular reads and writes to a specific \gls{MMIO} region.
While I/O instructions are always intercepted under \gls{SEV-ES}, regular reads and writes to the \gls{MMIO} region are not. 
Not intercepting regular reads and writes to \gls{MMIO} regions raises the issue that the \gls{HV} would not be able to infer which \gls{MMIO} operations should be performed. 
To work around this limitation, \gls{SEV-ES} uses the \textit{MMIO/NPF} sequence~\cite{GHCB2020}, in which the \gls{HV} sets a \texttt{Reserved} bit in the \gls{SLAT} table entries which correspond to the \gls{MMIO} region.
When the \gls{VM} accesses a page with a set \texttt{Reserved} bit, the CPU throws an \texttt{MMIO/NPF} exception, which invokes the \gls{VM}'s VC handler. 
The VC handler determines the type of access and the involved data values. 
Afterwards, it writes the \gls{GPA} and the length of the memory access to the \gls{GHCB} and executes a \texttt{VMGEXIT} in order to expose the information to the \gls{HV}.
Afterwards, if the \gls{VM} performed a read access from an \gls{MMIO} region, the \gls{HV} writes the value into the \gls{GHCB} \textit{scratch buffer}, and the \gls{VM}'s VC handler loads the value into the corresponding register.
In comparison, when writing to an \gls{MMIO} region, the \gls{VM} copies the value to the \gls{GHCB} scratch buffer, from where the \gls{HV} reads it.

At the time of writing, \gls{SEV-ES} is not officially supported in Linux, but the patches have been made public in mid-2019 and are in-review since early 2020.
In all iterations of the \gls{SEV-ES} Linux patch set, the VC handler does not verify the address and properties of the page with the set \texttt{Reserved} bit. 
This allows a malicious \gls{HV} to set the bit on any of the \gls{VM}'s pages, tricking the \gls{VM} to leak secret data or to read \gls{HV}-controlled values into registers.
Additionally, it allows the \gls{HV} to infer control-flow information as each memory access on a page with set \texttt{Reserved} bit traps into the \gls{HV}. 
This gives a malicious \gls{HV} fine-grained control on data exfiltration and injection by setting and removing the \texttt{Reserved} bit for the \gls{VM}'s memory pages.

\begin{figure}[htb]
    \begin{subfigure}{.7\columnwidth}
    \begin{center}
        \includegraphics[width=1.\columnwidth]{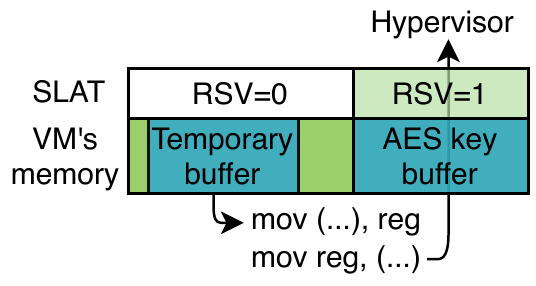}
        \caption{Data exfiltration from the \gls{VM}: 
        First, the \gls{HV} sets the \texttt{Reserved} bit on a page containing the \textit{AES key buffer}.
        Afterwards, writes to the page are exposed by the \gls{VM}'s VC handler, leaking the copied AES key.}
        \label{fig:rsv_exploits_a}
    \end{center}
    \end{subfigure}
    \begin{subfigure}{.7\columnwidth}
    \begin{center}
        \includegraphics[width=1.\textwidth]{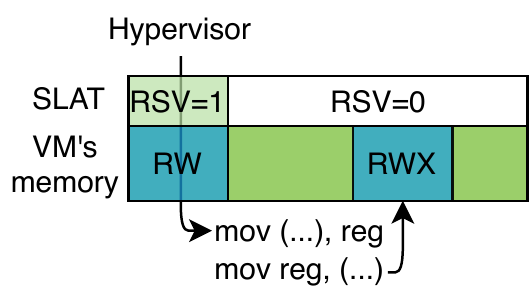}
        \caption{Code injection into the VM: 
        First, the \gls{HV} sets the \texttt{Reserved} bit on a page containing a Read-Write buffer with code. 
        This allows the \gls{HV} to intercept the read and to manipulate the information written into the executable buffer. }
        \label{fig:rsv_exploits_b}
    \end{center}
    \end{subfigure}
\end{figure}

When the Linux kernel derives an AES key, it first copies the key into a temporary buffer. 
This temporary buffer is then provided to the cipher implementation via the Kernel Crypto API, and the cipher implementation copies the buffer's content into an internal buffer. 
Figure~\ref{fig:rsv_exploits_a} shows an example for such a copy process, and how this allows a malicious \gls{HV} to extract data from the \gls{VM}. 
The \gls{HV}, aware of the buffer's location within the \gls{VM}'s address space, can set the \texttt{Reserved} bit on the page containing the \textit{AES key buffer}.
This would cause the write into the \textit{AES Key buffer} to throw an \texttt{MMIO/NPF} exception, causing the \gls{VM}'s VC handler to expose the data to be written to the \gls{HV}.

Further, the \gls{HV} can make use of a similar approach to gain code execution with an \gls{SEV-ES}-protected \gls{VM}. 
For the attack, the \gls{HV} determines a process within the \gls{VM} in which software copies data from a Read-Write buffer to another buffer, which is then marked as executable.
This process occurs naturally when the \gls{VM} loads a program for execution or when it compiles code \textit{just-in-time}.
Figure~\ref{fig:rsv_exploits_b} shows an example for such a process. 
Before the copy process, the \gls{HV} sets the \texttt{Reserved} bit for the page containing the Read-Write buffer. 
This allows the \gls{HV} to intercept the \gls{VM}'s read requests from the Read-Write buffer, and to provide malicious code and data.
Afterwards, the \gls{VM} copies the malicious code and data to the executable buffer, where the code is eventually executed by the assigned process within the \gls{VM}. 

While we showed the vulnerability using the example of the VC handler of the Linux kernel, we additionally verified that the vulnerability also exists in the \gls{OVMF} code base.
This makes the \gls{VM} also vulnerable to our attack at the very beginning of the boot sequence.
At this early stage, the \gls{VM}'s address space is not yet randomized, and most memory is marked as Read-Write-Executable.
These conditions allow the \gls{HV} to further simplify the attack and to gain code execution inside the \gls{VM} before the early Linux boot code has even begun execution.

\section{Code Execution via Guest Page Table Corruption}\label{sec:code_exec}
\glsreset{GPT}

In this section, we show how another approach to execute arbitrary code within an \gls{SEV} or \gls{SEV-ES}-protected \gls{VM}. 
To achieve this goal, we provide incorrect \texttt{CPUID} information when the \gls{VM} sets up its page tables at boot, allowing us to modify the \gls{VM}'s unencrypted stack.

\begin{figure}[htbp]
    \begin{lstlisting}[style=CStyle]
SYM_FUNC_START(get_sev_encryption_bit)
  ...
  movl $0x8000001f, eax
  cpuid
  bt $1, eax /* Check if SEV is available */
  jnc .Lno_sev
  ...
  movl ebx, eax
  andl $0x3f, eax /* Get C-bit location */
  jmp .Lsev_exit
\end{lstlisting}
    \caption{Code in Linux for checking the \gls{SEV} state. On Lines 3-6, the \gls{VM} checks if the CPU supports \gls{SEV} by executing \texttt{CPUID}.
    On Lines 8-9, the \gls{VM} retrieves the \textit{C-bit} position from the returned result.}
    \label{code:get_c_bit}
\end{figure}

Figure~\ref{code:get_c_bit} shows the code in the early Linux boot code  which checks whether \gls{SEV} is used and retrieves the location of the \texttt{C-bit} (Section~\ref{sec:amd_sev}) in a \gls{GPT} entry.
In Lines $3-6$, the \gls{VM} executes \texttt{CPUID} and checks whether \gls{SEV} is available by inspecting the first bit of the \texttt{EAX} register.
If \gls{SEV} is not available, the \gls{VM} returns.
Otherwise, the \gls{VM} extracts the position of the \texttt{C-bit} on Lines 8-9. 
Knowing the position of the \texttt{C-bit}, the \gls{VM} creates a bitmask which it applies to every entry in the \gls{GPT}. 
Afterwards, the \gls{VM} loads the \gls{GPT} by writing its address to the \texttt{CR3} register.

By intercepting the \texttt{CPUID} instruction in Line $4$, a malicious \gls{HV} can report \gls{SEV} as unavailable. 
The \gls{VM} would then mark all of its memory, including already encrypted pages, as unencrypted, causing a corruption of the \gls{GPT}. 
Although this corruption would mean that previously encrypted pages are not decrypted when read, the \gls{VM} would not crash immediately as instruction fetches and page table walks always interpret memory as encrypted~\cite{AMD2020}.
Thus, the \gls{VM} will continue execution until other accesses which do not decrypt the memory, such as data fetches, would cause a fault. 
This gives a malicious \gls{HV} sufficient time to overwrite the \gls{VM}'s stack, allowing to gain code execution inside the \gls{VM} via \glsfirst{ROP}.

In more detail, we target the function which setups and sets a new page table, \texttt{initialize\_identity\_map}.
The function initializes an identity mapping of memory for which a virtual address equals the physical address.
Afterwards, it sets the new \gls{GPT} by writing to the \texttt{CR3}, and returns.
This return will be followed by an immediate crash in case of a corrupted \gls{GPT}. 
The reason for the crash is that the saved \texttt{RIP} on the stack was encrypted, but is now accessed without decryption due to the corrupted \gls{GPT}.
This will cause an invalid instruction fetch, causing the \gls{VM} to crash.

However, a malicious \gls{HV} can avoid the crash by overwriting the \gls{VM}'s stack with suitable addresses before the return gets executed.
As both the \gls{HV} and the \gls{VM} access this memory without the \texttt{C-bit}, the \gls{VM} would see the same value as the \gls{HV}.
This enables the \gls{HV} to overwrite the \gls{VM}'s stack with return addresses, allowing to perform \gls{ROP}.

\begin{figure}[htb]
    \begin{center}
        \includegraphics[width=1.\columnwidth]{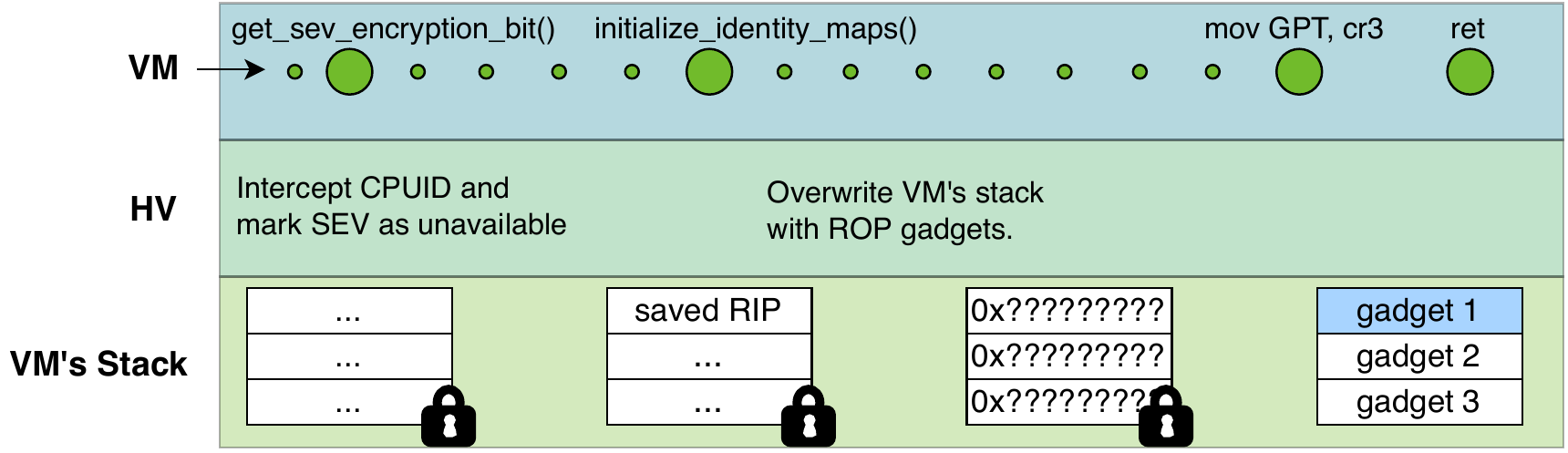}
    \end{center}
    \caption{Diagram for performing the \gls{GPT}-corruption attack. 
    First, the \gls{HV} provides invalid \gls{SEV} information and waits until the \gls{VM} enters the function \texttt{initialize\_identity\_maps()}. 
    Next, the \gls{HV} overwrites the \gls{VM}'s stack with a \gls{ROP}-chain which gets executed after the \gls{VM} sets the corrupted \gls{GPT} and returns.}
    \label{fig:rop}
\end{figure}

Figure~\ref{fig:rop} shows the detailed steps for performing the attack.
The top two lanes show the operations performed by the \gls{VM} and by the \gls{HV}, while the bottom lane shows the state of the \gls{VM}'s stack.
We again make use of page access tracking to track the \gls{VM}'s actions. 
When the \gls{VM} enters the function \texttt{get\_sev\_encryption\_bit()} and executes the \texttt{CPUID} instruction, the \gls{HV} intercepts the execution and reports that \gls{SEV} is not available.
Afterwards, the \gls{VM} calls the function \texttt{initialize\_identity\_maps()} by which it saves the \texttt{RIP} on the stack.
At this point, the \gls{HV} overwrites the \gls{VM}'s stack with the addresses of suitable \gls{ROP} gadgets in plaintext.
If the \gls{VM} would now access the data on the stack, it would attempt to decrypt the data written by the \gls{HV}, which would likely result in a crash. 
However, as part of creating the identity mapping, the \gls{VM} next sets the new, corrupted \gls{GPT} by writing to the \texttt{CR3} register.
Due to the corrupted \gls{GPT}, the \gls{VM} will no longer attempt to decrypt data on the stack, causing it to read the data exactly as previously written by the \gls{HV}. 
Therefore, after the \gls{VM} executes the return instruction, it executes the \gls{ROP}-chain injected by the \gls{HV}.
Using these steps, a malicious \gls{HV} is able to execute code inside an \gls{SEV}- or \gls{SEV-ES}-protected \gls{VM}.

\begin{figure}[htbp]
    \begin{lstlisting}[style=CStyle]
mov rcx, cr3 /* Change GPT */
lea 0x49401385(r8), rsp /* Control RSP */
add 0xd0248f72, eax
push rax /* Control RIP */
ret\end{lstlisting}
    \caption{\gls{ROP} gadget which sets the \gls{OVMF} \gls{GPT}, but also provides control over \texttt{RSP} and \texttt{RIP}.}
    \label{code:top_gadget}
\end{figure}

Despite being now able to execute code via \gls{ROP}, an attacker cannot directly steal any encrypted secrets from within the \gls{VM} as all of memory is marked as unencrypted.
To steal secrets or inject machine code, we need to map the corresponding pages as encrypted.
However, the \gls{VM} cannot modify its page table because \gls{SEV} requires the page table to be written encrypted~\cite{AMD2020}.
To still be able to continue exploitation, we make use of the page table created by \gls{OVMF}. 
The \gls{OVMF}'s \gls{GPT} contains two memory regions: an encrypted one for code and data, and an unencrypted one used for communication with the \gls{HV}.
However, simply performing the transition would crash the \gls{VM} as the \gls{ROP}-chain is written unencrypted, but the stack memory is located in the encrypted region of the \gls{OVMF}'s \gls{GPT}.
Therefore, after switching to the \gls{OVMF}'s \gls{GPT}, the \gls{ROP}-chain would be interpreted as encrypted memory and hold invalid addresses.
Thus, we require a gadget which updates \texttt{CR3} while allowing us to execute additional instructions to update the register state and avoid crashing the \gls{VM}.
To find such a gadget, we again make use of the \texttt{bzImage}, which is sufficiently large to contain a great selection of gadgets.
Figure~\ref{code:top_gadget} depicts an example for a gadget that we can use to continue exploitation.
With this gadget, we can first change to the \gls{OVMF}'s \gls{GPT} in Line $1$. 
In Line $2$, we position the stack into an unencrypted region by setting the \texttt{RSP}. 
Line $4$ allows us to push an arbitrary value onto the stack, which will then be used as \texttt{RIP}, allowing us to jump into another sequence of instructions.
At this point, we have both encrypted and unencrypted memory at our disposal, allowing us to inject instructions into the unencrypted region in order to execute arbitrary code as well as to read any secrets from the encrypted region.

\section{Evaluation}
\label{sec:evaluation}

In this section, we show the feasibility of our attack vectors by evaluating them in an \gls{SEV}- and \gls{SEV-ES}-protected environment.
We performed our evaluation on a system equipped with an AMD EPYC 3251 8-Core processor with $64$GB RAM running Debian $11$.
The host kernel was Linux \texttt{5.6.0} with the most recent patches for \gls{SEV-ES} support.
Additionally, we added the \texttt{RDTSC} and \texttt{TSC Ratio MSR} manipulation patches, the \gls{MMIO}-region-forging patch, the \gls{GPT}-corruption patch, and the patch from the SEVered framework~\cite{severedframework}.

We made use of \texttt{KVM} and \texttt{QEMU} in version \texttt{4.2.50} and two different \glspl{VM}.
The first \gls{VM} was running Ubuntu \texttt{18.04} with kernel version \texttt{4.15.0-88} and \gls{SEV} enabled, but without support for \gls{SEV-ES}.
The second \gls{VM} was running a custom built Linux kernel with version \texttt{5.8.0-rc4} and has the latest \gls{SEV-ES} patches applied.
Notably, we launched the \gls{SEV-ES} \gls{VM} with the \texttt{bzImage} directly provided to QEMU via the \texttt{-kernel} parameter.

We verified the attack vector based on untrusted \text{virtio} devices (Section~\ref{sec:steal}) on the \gls{SEV}-protected \gls{VM}.
We first tested that the \gls{HV} is able to provide arbitrary data to the \texttt{/dev/hwrng} device in the \gls{VM} provided by \texttt{virtio-rng}.
Further, we assured that the \gls{HV} is able to extract cryptographic keys and plaintext data by using the \texttt{virtio-crypto} device.
To ensure that the \gls{VM} used the device, we modified \texttt{KVM} to advertise the \texttt{AES-NI} extension as unavailable.
Lastly, we modified the default \gls{SEV} launch script to launch the \gls{VM} with the \texttt{virtio-crypto} device.
In order to extract the information communicated over the \texttt{virtio-crypto}, we modified the implementation in \texttt{QEMU} to print the encryption key and the data.
The selected test cases included \texttt{ip-xfrm}, \texttt{kcapi-enc}, and a custom kernel module in the \gls{VM} which used the Linux Crypto API to encrypt a secret message.
We used the two programs and the kernel module to encrypt secret data with \texttt{AES-CBC}.
In all three tests, we were able to extract the encryption keys and secret data from the \gls{VM} in plain text.

To evaluate our \texttt{RDTSC} manipulation technique through interception (Section~\ref{section:entropy_man}) and \texttt{TSC Ratio MSR} (Section~\ref{section:tscmsr}), we also used the \gls{SEV}-protected \gls{VM}.
As part of the verification, we used a kernel module in the \gls{VM} to print information about the kernel's probabilistic defenses.
The module printed the \gls{KASLR} virtual and physical offset, the offsets for the kernel memory regions and the kernel stack canaries of the first 15 running processes.
It is important to note that we used the module solely for evaluation purposes, and did not communicate with the module in any way during the execution of the attack itself.
To avoid maintenance of the \texttt{TSC Ratio MSR} among all cores, we pinned the \texttt{QEMU} process to a single core using the \texttt{taskset} command.
Additionally, we modified KVM to always zero the \texttt{TSC Offset} in the Virtual Machine Control Block.
For both the \texttt{RDTSC} and the \texttt{TSC Ratio MSR} manipulation, we performed $1000$ evaluation rounds.
In each round, we provided a value of zero for \texttt{RDTSC} whenever it was used as a source of entropy either through interception (Section~\ref{section:entropy_man}) or by modifying the \texttt{TSC Ratio MSR} (Section~\ref{section:tscmsr}).
After having performed the attack, we connected to the \gls{VM} via \texttt{SSH} to collect the information about the offsets.
At the end of each round, we rebooted the \gls{VM} to start the next round of the attack.
With both approaches, we were able to pin the value of the stack canaries of the first eleven processes for $1999$ out of the in total $2000$ runs.
The values in the one failing run and of further stack canaries have been likely influenced by interrupts and other events to the \gls{VM}, which caused entropy to be added to the \texttt{PRNG} entropy pool~\cite{muller2019documentation}.
Additionally, we were able to successfully pin all \gls{KASLR} offsets in $999$ runs, and only failed when the \gls{VM} is initially booted.
The failure to pin the offsets on the first boot is due to a bug in GRUB which does not correctly initialize the boot parameters structure when \gls{SEV} is used.

To prove that the failure during the initial boot is caused by incorrect initialization of the boot parameters structure, we additionally evaluated our attack on the \gls{SEV-ES}-protected \gls{VM} which does not use GRUB. 
For this, we modified the kernel image to directly print the \gls{KASLR} offsets once kernel initialization has finished.
We used the printed information to verify that we successfully pinned all offsets.
For both the \texttt{RDTSC} and the \texttt{TSC Ratio MSR} manipulation, we performed $20$ evaluation rounds.
At the end of each round, we killed the QEMU process and launched a new \gls{VM} instance using the same launch arguments.
With both approaches, we were able to successfully pin all \gls{KASLR} offsets in all $20$ runs for both \gls{SEV} and \gls{SEV-ES}.
These results confirm that our attack can also be applied at the initial boot of a \gls{VM}.

We then validated a \gls{PoC} based on forging an \gls{MMIO} region (Section~\ref{sec:mmio_npf}) over the encrypted memory of an \gls{SEV-ES}-protected \gls{VM}.
The \gls{PoC} featured a user space process which first allocated a \gls{RW} page and a \gls{RWX} page.
Next, the process wrote a sequence of \texttt{NOP} instructions followed by the \texttt{RET} instruction to the \gls{RW} page. 
Afterwards, it copied the contents of the \gls{RW} page to the the \gls{RWX} page and called into the RWX page.
To simplify the evaluation, the process reported the \gls{GPA} of the \gls{RW} page by issuing a \texttt{vmmcall} instruction, after which the \gls{HV} set the \texttt{Reserved} bit on the \gls{RW} page. 
On the \gls{HV}, we were able to intercept each read from the \gls{RW} page and to provide our own payload to be written to the \gls{RWX} page. 
Additionally, we verified with a similar \gls{PoC} exploit that the \gls{HV} can view writes to pages with set \texttt{Reserved} bit.

We also evaluated the \gls{GPT} corruption attack (Section~\ref{sec:code_exec}) on the \gls{SEV-ES}-protected \gls{VM}. 
During the evaluation, we overwrote the \gls{VM}'s stack with a \gls{ROP}-chain which executed a sequence of gadgets to write a fixed string to a fixed location in the \gls{VM}'s memory. 
This string could be read by the \gls{HV} due to the performed \gls{GPT} corruption.
$0.5$ seconds after overwriting the \gls{VM}'s stack, we read the \gls{VM}'s memory supposed to contain the fixed string from the \gls{HV} to determine if the attack was successful. 
Using this method, we performed 1000 evaluation rounds of the attack on the \gls{SEV-ES}-protected \gls{VM}.
After each round, we killed the \gls{VM} and launched a fresh instance to avoid the results being influenced by previous rounds. 
In all $1000$ rounds, we successfully wrote our string to the fixed location in the \gls{VM}'s memory.

\section{Discussion}
\label{sec:discussion}

Due to unintended behavior in the GRUB bootloader, the boot parameter structure of the \gls{SEV}-protected Ubuntu \gls{VM} is currently partially filled with random values when it is first booted.
This additional randomness prevents us from successfully pinning the \gls{KASLR} offsets (Section~\ref{sec:control}) at the initial launch of the \gls{SEV}-protected \gls{VM}, which uses GRUB.
However, as we showed in our evaluation (Section~\ref{sec:evaluation}), our technique works on every subsequent reboot of the Ubuntu \gls{VM}, which correctly initializes the boot parameters structure with zeroes.
Furthermore, our attack works on all launches of a barebone \gls{VM} with \gls{SEV-ES}, as GRUB is not used.
In this scenario, the \gls{OVMF}~\cite{ovmf} is responsible for populating the boot parameter structure before handing control to the Linux EFI boot stub~\cite{efi-boot-stub}.
While filling the boot parameters structure with random values could be interpreted as a defense mechanism against our approach, the Linux boot protocol requires the boot parameter structure to be initialized with zeros~\cite{linux-boot-protocol}.
This issue therefore only poses a temporary limitation of our attack for entropy manipulation.
We expect our approach to be fully compatible with all \gls{SEV} features as soon as the behavior of GRUB is adapted to fulfill the requirements of the Linux boot protocol under \gls{SEV}.

Using our approach for entropy manipulation (Section~\ref{sec:control}), we were able to pin the value of the first eleven stack canaries.
Although it should be possible to make the sequence of stack canaries deterministic over a longer period by manipulating the injection of interrupts, we did not further verify this possibility and leave this up to future work.

The entropy manipulation attack as described in this work applies to the current implementation of \gls{KASLR}.
However, Kristen Accardi recently proposed a modification to \gls{KASLR} to dynamically shuffle the kernel functions at boot time~\cite{edge2020finer}.
The corresponding code also relies on \texttt{kaslr\_get\_random\_long} to select a new position for each function.
Similar to previous examples, we would also in this scenario be able to control the source of entropy for the randomization function and thus be able to make the sequence of kernel functions deterministic.

One assumption we made for our entropy manipulation attack is that the \gls{HV} is able to access the \gls{VM}'s unencrypted kernel.
Also for the code execution attack, we make use of the \texttt{bzImage} that is being loaded, which provides a great amount of possible gadgets. 
However, the owner of the \gls{VM} may decide to encrypt the \texttt{/boot} partition with the \texttt{bzImage}, and have the decryption be performed either by the bootloader or by OVMF. 
However, rendering the kernel image inaccessible to the \gls{HV} would only be a temporary workaround to prevent our attack.
For entropy manipulation, we could for example use a modified version of the approach presented by Werner et al.~\cite{werner2019severest} to perform \gls{OS} fingerprinting.
Having determined the \gls{OS} version of the \gls{VM}, we would be able to download and inspect the respective kernel image.
Additionally, the code execution attack is likely also applicable to \gls{OVMF} since it also enables paging.
This would allow us to perform an attack similar to the one described, but early in the boot process, circumventing the issue of not being able to access the \texttt{bzImage}.

For our \gls{GPT}-corruption attack, our page tracking mechanism relies on specific conditions and would have to be adapted when the used triggers change within the \gls{VM}.  
To simplify determining when to overwrite the \gls{VM}'s stack, we can make use of other events such as the \texttt{VMEXIT\_CR[0-15]\_WRITE\_TRAP}. 
This trap causes a \texttt{VMEXIT} each time the \gls{VM} writes to any of the control registers~\cite{AMD2020}.
This allows us to detect any modification to the \texttt{CR0} and \texttt{CR3} registers, being able to precisely determine when to overwrite the \gls{VM}'s stack. 

The presented attacks in this paper exploit missing validation of results returned by the untrusted \gls{HV}.
An interesting question is whether these attack vectors can also be applied to \gls{SEV-SNP}, which is supported neither by any AMD CPU nor by Linux at the time of writing.
The \gls{SEV-SNP} specification~\cite{amd2020snpabi} describes two features which may effectively mitigate our attacks on entropy manipulation and \gls{GPT} corruption: \textit{CPUID Reporting} and \textit{CPUID Page}.
The \textit{CPUID Reporting} feature allows the \gls{VM} to query the CPU firmware for whether the \gls{HV} has reported \texttt{CPUID} features which are actually unsupported by the CPU.
However, this feature will not indicate if the \gls{HV} does not report features which are present~\cite{amd2020snpabi}.
Thus, \textit{CPUID Reporting} does not protect against the presented attacks since they all rely on reporting a feature being disabled.
In comparison, according to our understanding, the \textit{CPUID Page} feature is a page injected by the \gls{HV} when launching the \gls{VM}, and the content of the page is included into the attestation measurement.
One possible usage of the \textit{CPUID Page} feature is to have the \gls{HV} communicate all possible \texttt{CPUID} results once, and then have the VC handler query the structure when \texttt{CPUID} is intercepted.
This prevents the \gls{HV} from providing inconsistent information during the \gls{VM}'s execution.
However, both of these features need to be explicitly used by software in the \gls{VM}.

\section{Defenses}\label{sec:defenses}

While the issues described in this report also apply to traditional virtualized environments, they are of much higher relevance to \gls{SEV}-protected \glspl{VM} due to the stronger attacker model.
First, \gls{SEV} must blacklist security critical \texttt{virtio} devices such as \texttt{virtio-rng} and \texttt{virtio-crypto}, as they pose security risks for software which utilizes their respective interfaces.
This can be achieved by disabling \texttt{CONFIG\_HW\_RANDOM\_VIRTIO} and \texttt{CONFIG\_CRYPTO\_DEV\_VIRTIO} in the kernel build scripts provided by AMD~\cite{AMDSEV}.
We did create a pull request to apply this defense, which has been accepted and merged into the official repository~\cite{amdsevgit}.

To mitigate our attack for entropy manipulation (Section~\ref{sec:control}), it would be required to disallow the \gls{HV} to advertise \texttt{RDRAND} as unsupported.
Further, we suggest to make \texttt{RDRAND} a required feature for running \gls{SEV}-protected \glspl{VM}.
Based on a suggestion from Joerg Roedel (SUSE), we implemented a kernel patch to verify the cached CPU capability information in the early boot stage and in the kernel initialization phase.
The added validation is a sufficient countermeasure against the proposed attacks for pinning \gls{KASLR} offsets and stack canaries in the \gls{VM}'s Linux kernel.
The patch is included in the official \gls{SEV-ES} patch set~\cite{sev_es_patch_set} which is currently in review.

Another mitigation would be to prevent the \gls{HV} from manipulating the \gls{VM}'s view on the \texttt{TSC}, which has to be performed by the CPU firmware.
For this, a new \texttt{TSC} field could be added into the \gls{VMSA}, which is the structure that contains the vCPU's encrypted register state.
The firmware would be responsible for restoring the \texttt{TSC} register when the \gls{VM} is resumed and saving it when the \gls{VM} exits.

To mitigate the \gls{MMIO} forgery vulnerability (Section~\ref{sec:mmio_npf}), the \gls{VM}'s VC handler would have to check if the accessed page is unencrypted when receiving an \texttt{MMIO/NPF} exception. 
As \gls{MMIO} accesses will only happen on unencrypted pages, an \texttt{MMIO/NPF} exception raised from accessing an encrypted page indicates an attack. 
This behavior has been implemented for the Linux kernel by Joerg Roedel (SUSE) directly after disclosure of the attack~\cite{mmio_fix_joerg}.
A similar change would be necessary in \gls{OVMF} as well.

\begin{figure}[htbp]
    \begin{lstlisting}[style=CStyle]
mov GPT, cr3
cmpl 0xfff63d81, -10(rip)
je .access_is_ok
/* take appropriate actions */
.access_is_ok: /* test passed, then return */
ret\end{lstlisting}
    \caption{Code for detecting \gls{GPT} corruption. On Line $2$, the instruction compares the expected four bytes of its machine code with the four bytes in memory.
             If the test fails, the \gls{GPT} has been corrupted and the \gls{VM} can take action.}
    \label{code:access_check}
\end{figure}

Our last mitigation addresses the attack of corrupting the \gls{GPT} to gain code execution inside the \gls{VM} (Section~\ref{sec:code_exec}).
After corruption of the \gls{GPT}, the \gls{VM}'s instruction fetches remain encrypted while the \gls{VM}'s memory accesses now interpret memory as unencrypted.
This behavior can be used to create the mitigation shown in Figure~\ref{code:access_check}.
After the new \gls{GPT} is set on Line 1, the instruction on Line 2 reads the first four bytes of the current instruction and compares it with an immediate value which corresponds to the first four bytes of the machine code of the current instruction.
If the read bytes do not match the immediate, the \gls{GPT} has been corrupted, and the \gls{VM} can take appropriate measures. 
Otherwise, the \gls{VM} jumps to the label \texttt{.access\_is\_ok} and returns.
We verified that this test successfully detects our attack and protects against it.
However, the code running inside the \gls{VM} needs to be carefully analyzed to determine all places where the test needs to be applied.

\section{Related work}
\label{sec:related}

Checkoway et al.~\cite{checkoway2013iago} manipulated the return values of syscalls from the Linux kernel to cause protected applications to act against their own interests.
The authors showed that in order to prevent against kernel attacks, return values from the syscall interface must be sanitized.
Our work builds on the idea of Checkoway et al. and applies their approach to \gls{SEV}- and \gls{SEV-ES}-protected environments.

One of the first attacks on \gls{SEV} was presented by Hetzelt et al.~\cite{hetzelt2017security} and made use of the unprotected \gls{SLAT} table.
By remapping entries in the \gls{SLAT} to different pages in the physical memory, the authors were able to modify the control flow of an SSH server to allow an attacker to login without valid credentials.
The remapping approach was later extended by Morbitzer et al.~\cite{morbitzer2018severed}, who managed to extract the encrypted memory of a \gls{VM} in plaintext by making use of a service running inside the \gls{VM}.
The two attacks exploit the missing integrity protection of the \gls{SLAT}, and are similar to the exploitation of the \gls{MMIO} region forgery vulnerability discussed in Section~\ref{sec:mmio_npf}.
However, our approach offers fine-grained manipulation of memory accesses, and additionally allows for tracing the \gls{VM}'s execution since each memory access traps in the \gls{HV}.

Du et al.~\cite{du2017secure} were the first to reverse engineer the cipher mode used for encrypting the \gls{VM}'s memory.
This allowed them to patch instruction sequences of an SSH server running in the \gls{VM} in order to login without valid credentials.
The knowledge of the encryption algorithm was also used by Li et al.~\cite{li2019exploiting}, who exploited the I/O channel of the \gls{VM} to create an encryption oracle.
Afterwards, Wilke et al.~\cite{wilke2020sevurity} reverse engineered the cipher mode and tweakable functions on newer AMD CPUs.
They made use of their discovery by moving ciphertext blocks in the \gls{VM}'s memory in such a way that they were able to chain short sequences of code.
However, the two attacks are not applicable to recent generations of AMD CPUs where the tweakable functions are difficult to compute~\cite{wilke2020sevurity}.
In comparison, the \gls{MMIO} region forgery and GPT-corruption attacks do not require knowledge of the cipher mode and tweak values.
Thus, these two attacks are also applicable to recent generations of EPYC CPUs.

Werner et al.~\cite{werner2019severest} monitored general purpose registers of the \gls{VM} to extract confidential information such as encryption keys.
However, using \gls{SEV-ES}, the register state is protected, which prevents their attack.
In comparison, our approach of using the \texttt{virtio-crypto} devices allows us to extract secret keys even in \gls{SEV-ES} protected \glspl{VM}.
Furthermore, our \gls{MMIO} region forgery attack can be used to leak the address and data values of a memory write with \gls{SEV-ES}, which similarly can be employed to steal encryption keys.
Another advantage lays in our GPT-corruption approach, which can be used to extract or inject arbitrary data and also works on \gls{SEV-ES}.

Werner et al.~\cite{werner2019severest} also presented a second attack, which allows to fingerprint applications running in the protected \gls{VM} by using \gls{IBS} even under \gls{SEV-ES}.
Morbitzer et al.~\cite{morbitzer2019extracting} took a different approach, and located confidential information in a \gls{VM}'s encrypted memory by purposely removing page access rights in the \gls{SLAT} table and analyzing the \gls{VM}'s access pattern.
To extract the information, they require a method to extract the located data.
Also Buhren et al.~\cite{buhren2018detectability} analyzed access patterns of \glspl{VM}.
By using machine learning, they were able to detect which syscalls applications in the \gls{VM} were issuing.
Our work on fixing the \gls{KASLR} offsets can be seen as a great addition to those contributions.
For example, disabling \gls{KASLR} could simplify locating secret data within the \gls{VM}'s encrypted memory, or to identify syscalls.
Additionally, our \gls{MMIO} region forgery attack would allow to extract or modify data which has been located within the \gls{VM}'s memory using the approach of Morbitzer et al.~\cite{morbitzer2019extracting}.

\section{Conclusion}

In this work, we showed the security implications of introducing a stricter attacker model into a complex code base for which some interfaces are no longer trusted.
We emphasized the risk of this scenario by showing how a malicious \gls{HV} can 1) extract cryptographic keys through virtual devices, 2) disable probabilistic software defenses, 3) intercept regular memory accesses and 4) gain code execution inside the encrypted \gls{VM}.

We showed how a malicious \gls{HV} can make use of \texttt{virtio} devices to extract secret data such as encryption keys or plaintext data from the \gls{VM}'s Kernel Crypto API.
Further, we utilized security critical \texttt{virtio} devices to control the entropy for software running in the encrypted \gls{VM}.
Also, we presented an approach for a malicious \gls{HV} to influence the generation of random numbers in an \gls{SEV}- and \gls{SEV-ES}-protected \gls{VM} by intercepting CPU instructions.
This enabled us to reduce the entropy of probabilistic defenses such as \gls{KASLR} and stack canaries in the \gls{VM}'s kernel.
Afterwards, we demonstrated how missing validation in the handling of \texttt{MMIO/NPF} events can lead to data exfiltration and code injection for an \gls{SEV-ES}-protected \gls{VM}.
Finally, we presented an approach to trick an \gls{SEV}- or \gls{SEV-ES} \gls{VM} to decrypt its stack which leads to code execution via \gls{ROP}.

While these attacks can be applied to all virtualized environments, they are especially critical for environments such as \gls{SEV} and \gls{SEV-ES}, in which the \gls{HV} is considered untrusted.
We showed this at the example of the Linux kernel, which is making use of \gls{HV}-controlled interfaces for critical security features.
Our work reveals that software running within an \gls{SEV}- or \gls{SEV-ES}-protected \gls{VM} must not trust any input from the \gls{HV} and carefully verify all external data.

\section{Acknowledgments} 
This work has been funded by the Fraunhofer Cluster of Excellence ``Cognitive Internet Technologies''\footnote{\url{https://www.cit.fraunhofer.de}}.   

We would like to thank David Kaplan and Tom Lendacky from AMD, as well as Joerg Roedel from SUSE for the quick responses and fixes.

\bibliographystyle{ACM-Reference-Format}
\bibliography{bibliography}

\end{document}